\documentclass[12pt]{article}

\usepackage{amssymb}
\usepackage{amsmath}

\begin{document}
\title{Towards chaos criterion in quantum field theory.}
\author{V.I. Kuvshinov$^1$, A.V. Kuzmin$^2$ \\
$1$-Institute of Physics, 220072 Belarus, Minsk, Scorina,68.\\
                    Tel. 375-172-84-16-28, fax: 375-172-84-08-79.\\
                    E-mail: kuvshino@dragon.bas-net.by \\
$2$-Institute of Physics, 220072 Belarus, Minsk, Scorina,68.\\
                    E-mail: avkuzmin@dragon.bas-net.by}

\date{}
\maketitle

\begin{abstract}
Chaos criterion for quantum field theory is proposed. Its correspondence with
classical chaos criterion in semi-classical regime is shown. It is demonstrated for
real scalar field that proposed chaos criterion can be used to investigate stability
of classical solutions of field equations.
\end{abstract}


Phenomenon of chaos attracts much attention in various fields of physics. Originally
it was associated with problems of classical mechanics and statistical physics.
Substantiation of statistical mechanics initiated intensive study of chaos and
uncovered its basic properties mainly in classical mechanics. One of the main results
in this direction was a creation of KAM theory and understanding of the phase space
structure of Hamiltonian systems \cite{Kolmogorov},\cite{Arnold}. It was clarified
that the root of chaos was local instability of dynamical system \cite{Krylov}. Local
instability leads to mixing of trajectories in phase space and thus to non-regular
behavior of the system and chaos \cite{Lihtenberg}-\cite{Zaslavski}. Significant
property of chaos is its prevalence in various natural phenomena. It explains a large
number of works in this field.
\par
Large progress is achieved in understanding of chaos in semi-classical regime of
quantum mechanics via analysis of the spectral properties of the system
\cite{Zaslavski},\cite{Robnic}. Semi-classical restrictions are important, because a
large number of energy levels in small energy interval is needed to provide a certain
statistics \cite{Robnic2}.
\par
Investigation of the stability of classical field solutions faces difficulties caused
by infinite number of degrees of freedom. That is why authors often restrict their
consideration by investigation of some model field configurations \cite{Kawabe}.
\par
There are papers devoted chaos in quantum field theory \cite{Q2}. But there is no
generally recognized definition of chaos for quantum fields \cite{Bunakov}. This fact
restricts use of chaos theory in field of elementary particle physics. At the same
time it is well known that field equations  of all four types of fundamental
interactions have chaotic solutions \cite{Int4} and high energy physics reveales the
phenomenon of intermittency \cite{Intermittency}. The aim of this work is to propose
chaos criterion applied for both quantum mechanical and quantum field systems.
\par
The paper is structured as follows. At first we generalize classical Toda criterion of
local instability \cite{Toda}-\cite{Ukraine} for systems with any finite number of
degrees of freedom. Then using analogy with statistical mechanics we formulate chaos
criterion for quantum mechanics and quantum field theory. After that we apply
formulated chaos criterion to quantum mechanical systems in semi-classical
approximation and check its accordance with generalized Toda criterion for the
corresponding classical system (correspondence principle). And finally we apply
formulated chaos criterion to analyze $\varphi^{4}$-theory of real scalar field.

Toda criterion of local instability for classical mechanical systems was at first
formulated in \cite{Toda}. It was reformulated for Hamiltonian systems with two
degrees of freedom by Salasnich \cite{Salasnich}. Local instability of classical
conservative Hamiltonian system with finite number of freedoms and  finite (at finite
energy) available phase space volume leads to mixing, destruction of the first
integrals of motion and all that one calls chaos \cite{Lihtenberg}-\cite{Zaslavski}.
Agreement between Toda criterion and criterion of classical chaos based on KAM theory
and conception of nonlinear resonance was checked on the particular example in
\cite{we3}.
\par
Consider classical Hamiltonian system with any finite number of degrees of freedom
\begin{equation}
H=\frac{1}{2}\vec{p}^{2}+V(\vec{q})\quad,\quad
\vec{p}=(p_{1,\ldots,}p_{N})\;;\;\vec{q}=(q_{1,\ldots ,}q_{N}), \quad N>1.
\end{equation}
Behavior of the classical system is locally unstable if distance between two
neighboring trajectories grows exponentially with time in some region of the phase
space.
\par
 Consider small region $\Omega$ of the phase space near the point
$(\vec{q}_{0},\vec{p}_{0})$. Suppose that there are two classical trajectories
$(\vec{q}^{(1)}(t),\vec{p} ^{(1)}(t))$ and $(\vec{q}^{(2)} (t),\vec{p}^{(2)}(t))$ in
$\Omega$. Then deviations of trajectories are
\begin{equation}
\smallskip\delta\vec{q}(t)=\vec{q}^{(1)}
(t)-\vec{q}^{(2)}(t) \quad , \quad \delta\vec{p}(t)=\vec{p}^{(1)}(t)-\vec
{p}^{(2)}(t).
\end{equation}
Evolution of these deviations in $\Omega$ is governed by the following linearized
Hamilton equations
\begin{equation}\label{lineareq}
\frac{d}{dt}\left(
\begin{array}
[c]{l}%
\delta\vec{q}\\
\delta\vec{p}%
\end{array}
\right)  = G  \left(
\begin{array}
[c]{l}%
\delta\vec{q}\\
\delta\vec{p}%
\end{array}
\right) \quad , \quad G\equiv\left(
\begin{array}
[c]{cc}%
0 & I\\
-\Sigma & 0
\end{array}
\right).
\end{equation}
Here $I$ is the $N \times N$ identity matrix, $G$ is a stability matrix and matrix $\Sigma$ is%
\begin{equation}\label{Sigma}
\Sigma\equiv\left(  \frac{\partial^{2}V}{\partial q_{i}\partial q_{j}}\left|
_{\vec{q}_{0}}\right.  \right).
\end{equation}
Matrix $\Sigma$ and stability matrix $G$ are functions of the point $\vec{q}_{0}$ of
configuration space of the system. Solution of the equations (\ref{lineareq}) valid in
$\Omega$ has the following form
\begin{equation}\label{solution}
\left(
\begin{array}
[c]{l}%
\delta\vec{q}(t)\\
\delta\vec{p}(t)
\end{array}
\right)  =\sum_{i=1}^{2N}C_{i}\exp\left\{  \lambda_{i}t\right\} \left(
\begin{array}
[c]{l}%
\delta\vec{q}(0)\\
\delta\vec{p}(0)
\end{array}
\right).
\end{equation}
Here $\lambda_{i}=\lambda_{i}(\vec{q}_{0})$ are eigenvalues of the stability matrix
$G$. And $\{C_{i}\}$ is a full set of projectors. From (\ref{solution}) it is seen:
\par
a) If there is $i$ such as $\operatorname{Re}\lambda _{i}>0$ then the distance between
neighboring trajectories grows exponentially with time and motion is locally unstable.
According Liouville's theorem stretching of phase space flow in one direction
($\operatorname{Re} \lambda_{i} >0$) is accompanied by its compression in other
direction (directions) in order to keep phase space volume constant. That means the
existence of $\operatorname{Re} \lambda_{j} < 0$. Thus for local instability of motion
we can demand existence of $\operatorname{Re}\lambda _{k} \neq 0$.
\par
b) If for any $i=\overline{1,2N}\quad$ $\operatorname{Re}\lambda_{i}=0$ \ then there
is no local instability and the motion is regular.
\par
It is easy to see that $G^{2}=\operatorname{diag}(-\Sigma, -\Sigma)$.Therefore if
$(-\xi_{i})\;,\;i=\overline{1,N}$ are eigenvalues of the matrix $(-\Sigma)$ then
\begin{equation}
(-\xi_{i})=\lambda_{i}^{2}\;,\;i=\overline{1,N} \quad ,
 \quad \lambda_{i}^{2}=\lambda_{i+N}^{2}\;,\;i=\overline{1,N}.%
\end{equation}
Thus without loss of generality we can imply that $\operatorname{Re}\lambda_{i}\geq0$.
Notice that
\begin{equation}\label{123}
\xi_{i}=-\lambda_{i}^{2}=(\operatorname{Im}\lambda_{i})^{2}-
(\operatorname{Re}%
\lambda_{i})^{2}- 2i\operatorname{Im}\lambda_{i}\operatorname{Re}\lambda
_{i}\;,\;i=\overline{1,N}.%
\end{equation}
Since matrix $\Sigma$ is real and symmetric its eigenvalues $\{\xi _{i}\} ,\quad
i=\overline{1,N}$ are real. Therefore (\ref{123}) leads to the expression
\begin{equation}
\operatorname{Im}\lambda_{i}\operatorname{Re}\lambda_{i}=0\qquad
 \forall
i=\overline{1,N}.%
\end{equation}
Thus any eigenvalue of the stability matrix $G$ is real or pure imaginary or equals
zero. Therefore the generalized Toda criterion for classical Hamiltonian systems with
any finite number of freedoms can be formulated as follows:
\par
a) If $\xi_{i}\geq0 \quad , \quad \forall i=\overline{1,N}$ \ then behavior of the
system is regular near the point $\vec{q}_{0}$.
\par
b) If $\exists i=\overline{1,N}:\xi_{i}<0$ then behavior of the system is locally
unstable near the point $\vec{q}_{0}$.
\par
If one of these conditions holds in some region of the configuration space then the
motion is stable or chaotic respectively in this region. These results for systems
with two degrees of freedom coincide with ones obtained in \cite{Salasnich}.
\par
Now we give some qualitative arguments which bring us to formulation of chaos
criterion in quantum mechanics and quantum field theory. From statistical mechanics
and ergodic theory it is known that chaos in classical systems is a consequence of the
property of mixing \cite{Lihtenberg}-\cite{Zaslavski}. Mixing means rapid
(exponential) decrease of correlation function with time \cite{Zaslavski}. In other
words, if correlation function exponentially decreases than the corresponding motion
is chaotic, if it oscillates or is constant then the motion is regular \cite{Shuster}.
We expand criterion of this type for quantum field systems.  All stated bellow remains
valid for quantum mechanics, since mathematical description via path integrals is the
same.
\par
For field systems the analogue of classical correlation function is two-point
connected Green function
\begin{equation}\label{green}
G_{ik}(x,y)= -\frac{\delta^{2}W[\vec{J}]}{\delta J_{i}(x)
 \delta J_{k}(y)}\left.\right|_{\vec{J}=0}.
\end{equation}
Here $W[\vec{J}]$ is generating functional of connected Green functions, $\vec{J}$ are
the sources of the fields, $x$, $y$ are 4-vectors of space-time coordinates.
\par
Thus we formulate chaos criterion for quantum field theory and quantum mechanics in
the following form:
\par
a) If two-point Green function (\ref{green}) exponentially goes to zero when the
distance between its arguments goes to infinity then system is chaotic.
\par
b) If it oscillates or remains constant in this limit then we have regular behavior of
quantum system.
\par
To check the agreement between generalized Toda criterion and formulated quantum chaos
criterion in semi-classical limit we shall calculate two-point Green function in
semi-classical approximation of quantum mechanics. Generating functional is
\begin{equation}
Z[\vec{J}]= \int D\vec{q} \exp{\{i \int^{+\infty}_{- \infty} dt \left[
\frac{1}{2}\dot{\vec{q}}^{2} - V(\vec{q}) + \vec{J}^{T}\vec{q} \right] \}}.
\end{equation}
Here index $T$ denotes transposition. Consider certain solution of classical equations
of motion $\vec{q}_{0}(t)$. Introduce new variable describing deviations from the
classical trajectory $ \delta\vec{q}(t)=\vec{q}-\vec{q}_{0}(t)$, then under
semi-classical approximation
\begin{equation}\label{GF}
Z[\vec{J}] = \exp{\{i S_{0}[\vec{J}]\}} \int D\delta\vec{q} \exp{\{i
\int^{+\infty}_{-\infty} dt \left[\frac{1}{2}\delta\dot{\vec{q}}^{2} -
\frac{1}{2}\delta\vec{q}^{T} \Sigma \delta\vec{q} + \vec{J}^{T}\delta\vec{q} \right]
\}}.
\end{equation}
Here classical action
\begin{equation}
S_{0}[\vec{J}] = \int^{+\infty}_{-\infty} dt \left[\frac{1}{2} \dot{\vec{q}}_{0}^{2} -
V\left(\vec{q}_{0}(t)\right) + \vec{J}^{T}\vec{q}_{0} \right],
\end{equation}
and matrix $\Sigma$ defined in (\ref{Sigma}) is real and symmetric. Therefore there
exists orthogonal transformation $O$ reducing $\Sigma$ to diagonal form. For
simplicity we suppose that $\xi_{i}= \xi_{i}(\vec{q}_{0}(t)), \quad i=\overline{1,N}$
do not depend on time and remain constant on the classical trajectory. After
orthogonal transformation given by $\delta \vec{q} = O \vec{x},\quad \vec{J}^{T} =
\vec{\eta}^{T} O^{T}, \quad O^{T} \Sigma O = \operatorname{diag}(\xi_{1}, \ldots ,
\xi_{N}); \quad O^{T}O = OO^{T} = I$ we obtain
$$
Z[\vec{\eta}] = L \exp{\{i S_{0}[\vec{\eta}]\}} \times
$$
\begin{equation}\label{MIS}
 \int D\vec{x} \exp{\{i\int_{-\infty}^{+\infty} dt \left( \frac{1}{2}\dot{\vec{x}}^{2}
 - \frac{1}{2}\vec{x}^{T} \operatorname{diag(\xi_{1}, \ldots , \xi_{N})\vec{x} +
 \vec{\eta}^{T}\vec{x}} \right) \}},
\end{equation}
here $L$ denotes Jacobian of the orthogonal transformation. After analytical extension
of generating functional (\ref{MIS}) into Euclidian space and path integration we get
$$
Z_{E}[\vec{\eta}] = N \exp{(-S^{E}_{0}[\vec{\eta}])}\times
$$
\begin{equation}\label{1}
\times \prod^{N}_{i=1}\exp{\frac{1}{2} \int^{+\infty}_{-\infty} d \tau_{1}d\tau_{2}
\eta_{i}(\tau_{1})\left[ \delta(\tau_{1}-\tau_{2})  \left(
-\frac{d^{2}}{d\tau_{2}^{2}} + \xi_{i} \right) \right]^{-1} \eta_{i}(\tau_{2})}.
\end{equation}
Here and further there is no sum by $i$, N is a normalization factor. Classical
Euclidian action $S^{E}_{0}[\vec{\eta}]$ is a linear functional of the sources
$\vec{\eta}$. Inverse operator
\begin{equation}
\Delta_{i}(\tau_{1},\tau_{2}) = \left[\delta(\tau_{1} - \tau_{2}) \left(
-\frac{d^{2}}{d\tau_{2}^{2}} + \xi_{i} \right)
 \right]^{-1}
\end{equation}
has to satisfy the following equation
\begin{equation}
\left( -\frac{d^{2}}{d \tau_{1}^{2}} + \xi_{i} \right) \Delta(\tau_{1},\tau_{2}) =
\delta(\tau_{1} - \tau_{2}).
\end{equation}
The solution is
\begin{equation}
\Delta_{i}(\tau_{1},\tau_{2}) = \frac{1}{2\pi} \int d\omega
\frac{e^{i\omega(\tau_{1}-\tau_{2})}}{\omega^{2} + \xi_{i}}.
\end{equation}
Euclidian connected two-point Green function equals $\Delta_{i}(\tau_{1},\tau_{2})$.
Its analytical extension to real (physical) time is
\begin{equation}\label{Green}
G_{i}(t_{1},t_{2})= \frac{i}{2\pi} \int d\varpi \frac{e^{i\varpi(t_{1} -
t_{2})}}{\varpi^{2} + \lambda_{i}^{2}},
\end{equation}
here $\tau \rightarrow it$,$\omega \rightarrow -i\varpi$. Green function is defined up
to any solution of corresponding homogeneous equation. We use this freedom to make
Green function finite in the limit $(t_{1}-t_{2})\rightarrow +\infty$ for real
$\lambda_{i}$ and to obtain single formula for any $\lambda_{i}$ (both real and
imaginary). Thus two-point connected Green function (\ref{Green}) can be represented
in the form
\begin{equation}\label{MainGreen2}
 G_{i}(t_{1},t_{2})=\frac{i}{2}\operatorname{Re}\left( \frac{e^{-\lambda_{i}(t_{1} -
 t_{2})}}{\lambda_{i}} \right), \quad t_{1}>t_{2}.
\end{equation}
From the expression (\ref{MainGreen2}) it is seen
\par
a) If classical motion is locally unstable (chaotic) then according Toda criterion
there is real eigenvalue $\lambda_{i}$. Therefore Green function (\ref{MainGreen2})
exponentially goes to zero for some $i$ when $(t_{1}-t_{2})\rightarrow +\infty$.
Opposite is also true. If Green function (\ref{MainGreen2}) exponentially goes to zero
under the condition $(t_{1}-t_{2})\rightarrow +\infty$ for some $i$, then there exists
real eigenvalue of the stability matrix and thus classical motion is locally unstable.
\par
b) If all eigenvalues of the stability matrix $G$ are pure imaginary, that corresponds
classically stable motion, then in the limit $(t_{1}-t_{2})\rightarrow +\infty$ Green
function (\ref{MainGreen2}) oscillates as a sine. Opposite is also true. If for any
$i$ Green functions oscillate in the limit $(t_{1}-t_{2})\rightarrow +\infty$ then
$\{\lambda_{i}\}$ are pure imaginary for any $i$ and classical motion is stable and
regular.
\par
Thus we have demonstrated that proposed quantum chaos criterion coincides with Toda
criterion in the semi-classical limit (corresponding principle).
\par
One of possible applications of proposed chaos criterion in field theory is an
investigation of the stability of classical solutions with respect to small
perturbations of initial conditions. Of course, this does not directly imply chaos,
but advances us to it. To study the stability of certain classical solution of field
equations one has to calculate (for instance, using one loop approximation) two-point
Green function in the vicinity of considered classical solution.
\par
To demonstrate this, consider real scalar $\varphi^{4}$-field
\begin{equation}\label{L}
L= \frac{1}{2} \left( \partial_{\mu} \varphi \right)^{2} - \frac{1}{2}m^{2}
\varphi^{2} - \frac{\lambda}{4!} \varphi^{4}.
\end{equation}
Here $\lambda>0$ is a coupling constant, $m^{2}$ is some parameter which can be larger
or less then zero. In both cases $\varphi=0$ is a solution of field equations.
Asymptotic of two-point Green function calculated in the vicinity of the classical
solution $\varphi=0$ in the zero order of perturbation theory is
\begin{equation}\label{Lastf}
G(x,y)_{\widetilde{\rho \rightarrow \infty}} \rho^{-\frac{1}{2}}e^{im\sqrt{\rho}}.
\end{equation}
Here $\rho=(x-y)^{2}$ and we accept that 4-vector $x-y$ is inside the light cone
$(x^{0}-y^{0})>0$, in other words $\rho>0$. We can study the stability of considered
solution with respect to small perturbations. Expression (\ref{Lastf}) shows that we
have two different cases
\par
a) Green function oscillates and slowly (non-exponentially) goes to zero when $\rho
\rightarrow \infty$. According proposed chaos criterion considered solution is stable.
Indeed, from (\ref{Lastf}) it follows that parameter $m$ is real in this case.
Therefore $\varphi=0$ is a stable vacuum state.
\par
b) Green function exponentially goes to zero in the limit $\rho \rightarrow \infty$.
From proposed chaos criterion it follows that $\varphi=0$ is an unstable solution.
That is true since from (\ref{Lastf}) one can see that parameter $m$ has to be pure
imaginary. It is known that in this case state $\varphi=0$ becomes unstable, two new
stable vacuums are appeared and we obtain spontaneous symmetry breakdown \cite{Higgs}.
\par
Thus in this paper we have generalized Toda criterion for the Hamiltonian systems with
any finite number of degrees of freedom. Basing on the formal analogy between
statistical mechanics and quantum field theory we proposed chaos criterion for quantum
mechanical and quantum field systems. We have demonstrated that proposed chaos
criterion corresponds to generalized Toda criterion in semi-classical limit of quantum
mechanics in the case when Lyapunov exponents do not depend on time. For real scalar
$\varphi^{4}$-field we analysed the stability of vacuum state and showed  that
spontaneous symmetry breakdown and degeneration of vacuum state can be regarded as
signatures of quantum chaos.

\end{document}